\documentstyle[preprint,aps,psfig,epsf]{revtex}

\begin{document}
\title{Off equilibrium magnetic properties in a model of repulsive particles 
for vortices in superconductors}
\author{Mario Nicodemi and Henrik Jeldtoft Jensen}
\pagestyle{myheadings}
\address{Department of Mathematics, Imperial College, 180 Queen's Gate, 
London SW7 2BZ, UK}

\maketitle
\date{\today}
\begin{abstract}
We study the properties of a simple lattice model of repulsive particles 
diffusing in a pinning landscape. The behaviour of the model is very similar 
to the observed physics of vortices in superconductors. We compare and discuss
the equilibrium phase diagram, creep dynamics, 
the Bean critical state profiles, 
hysteresis of magnetisation loops (including the second peak feature),  
and, in particular, ``aging'' in relaxations. 
\end{abstract}
\vskip1pc


Important dynamical phenomena ranging from slow relaxations or hysteresis, 
to the anomalous ``second peak'' in magnetisation loops, 
are found  in vortex physics of many different superconductors within 
a broad range of material parameters. This observation suggests that 
some basic general mechanisms are responsible for the observed phenomenology 
\cite{blatter,Brandt,Yeshurun,cohen} and that 
schematic models from statistical mechanics can be 
successfully used to describe vortex matter 
\cite{blatter,Brandt,Yeshurun,cohen,Nelson,SMT,bassler1}. 

We consider here a simple statistical mechanics model 
that appears to reproduce a very wide range of properties of vortices, 
ranging from dynamical behaviours to phase transitions. The model is  
 an extension of a Multiple Occupancy 
cellular-automaton-like Model recently introduced by Bassler and Paczuski (BP) 
\cite{bassler1} to study vortex dynamics at coarse grained level. We 
introduce the vortex Hamiltonian
in order to be able to consider non-zero temperature effects in a consistent 
way and study them by Monte Carlo (MC) and replica theory methods. 
Our extension of the  BP model also 
limits the occupancy of the individual lattice sites to correctly
take into account the finiteness of the upper critical field.  This point is
of crucial importance for the phenomenological predictions of the model.
This leads us to a Restricted Occupancy Model (ROM).

We find that even the two dimensional version of the model 
is able to qualitatively reproduce many features similar to those 
observed in real superconducting samples, including a 
reentrant equilibrium phase diagram,  creep dynamics, 
hysteresis of  magnetisation loops, ``second peak'', and others. 
Here, in particular, we describe its off equilibrium magnetic properties. 
The model, simple and thus tractable,
nevertheless appears to capture significant aspects 
of the essential physics and help to establish a simple unified 
reference frame. 


{\it The model --} A detailed description of the
interaction potential, $U({\vec r})$, 
between vortices depends on the considered region in the
temperature-magnetic field ($T-H$) plane. For instance, at low field the
London approximation can be used to derive two body potentials
\cite{Brandt}, whereas at elevated fields other
approximations, such as the lowest Landau level approximation,
may become relevant (see eg. \cite{moore}). 
Like in the BP model we consider here a coarse grained
lattice version of an interacting vortex system, with a coarse
graining length scale, $l_0$, of the order of the natural screening 
length of the problem (typically, the magnetic penetration
length $\lambda$). 
After coarse graining, the original interaction potential, $U$, is reduced to 
an effective Hamiltonian coupling $A$. 
In this way a drastic reduction of degrees of freedom is accomplished 
and the resulting schematic effective model can be more easily dealt with. 
The price to pay is the the loss of information on scales smaller than 
$l_0$. 
However, some general features of the system behaviour can survive 
at the level of the coarse grained description. 
In the above perspective, below we only consider the 
essential properties of vortices interaction, i.e., 
a mutual repulsion amongst vortices together
with a spatially inhomogeneous pinning interaction. 
The present description can be, of course, refined by reducing the 
value of $l_0$. We consider the Hamiltonian: 
\begin{equation}
{\cal H}= \frac{1}{2} \sum_{ij} n_i A_{ij} n_j 
-\frac{1}{2} \sum_i A_{ii} n_i - \sum_i A^p_i n_i
\label{H}
\end{equation}
In eq.(\ref{H}), 
$n_i\in\{0,...,N_{c2}\}$ is an integer occupancy 
variable equal to the number of particles on site $i$. 
The parameter $N_{c2}$ importantly
bounds the particle density per site below a critical value and  represents
 the upper critical field $B_{c2}$ in 
type II superconductors. 
Particles also have a ``charge'' $s_i=\pm 1$ and 
neighbouring particles with opposite ``charge'' annihilate. 
The first term in eq.(\ref{H}) represents the repulsion between 
the particles \cite{Brandt}. Since the coarse graining length is taken to be
of order $\lambda$
we choose a finite range potential: $A_{ii}=A_0$; $A_{ij}=A_1$ 
if $i$ and $j$ are nearest neighbours;  
$A_{ij}=0$ for all others couples of sites. 
The second term in eq.(\ref{H}) just normalises the particle self-interaction
energy. The third term corresponds to a random pinning potential, with a 
given distribution $P(A^p)$, acting on a fraction $p$ of lattice sites
(below we use $p=1/2$). 
For simplicity we choose a delta-distributed random pinning: 
$P(A^p)=(1-p)\delta(A^p)+p\delta(A^p-A^p_0)$. 
To control the overall system ``charge density'' 
we can add a chemical potential term 
$-\mu\sum_i S_i$ to the above Hamiltonian ($S_i=s_in_i$). 
The parameters entering the model can be qualitatively related to material 
parameters of superconductors. 
The inter-vortex coupling $A_0$ sets the energy scale. 
The ratio $\kappa^*=A_1/A_0$ can be related to the 
Ginzburg-Landau parameter $\kappa=\lambda/\xi$ \cite{nota1} and, 
in general, is expected to be an increasing function of $\kappa$. 
The last parameter $A^p$ is a fraction of $A_0$. 

\begin{figure}[ht]
$~$\vspace{-2.7cm}
\centerline{
\hspace{-2.5cm}\psfig{figure=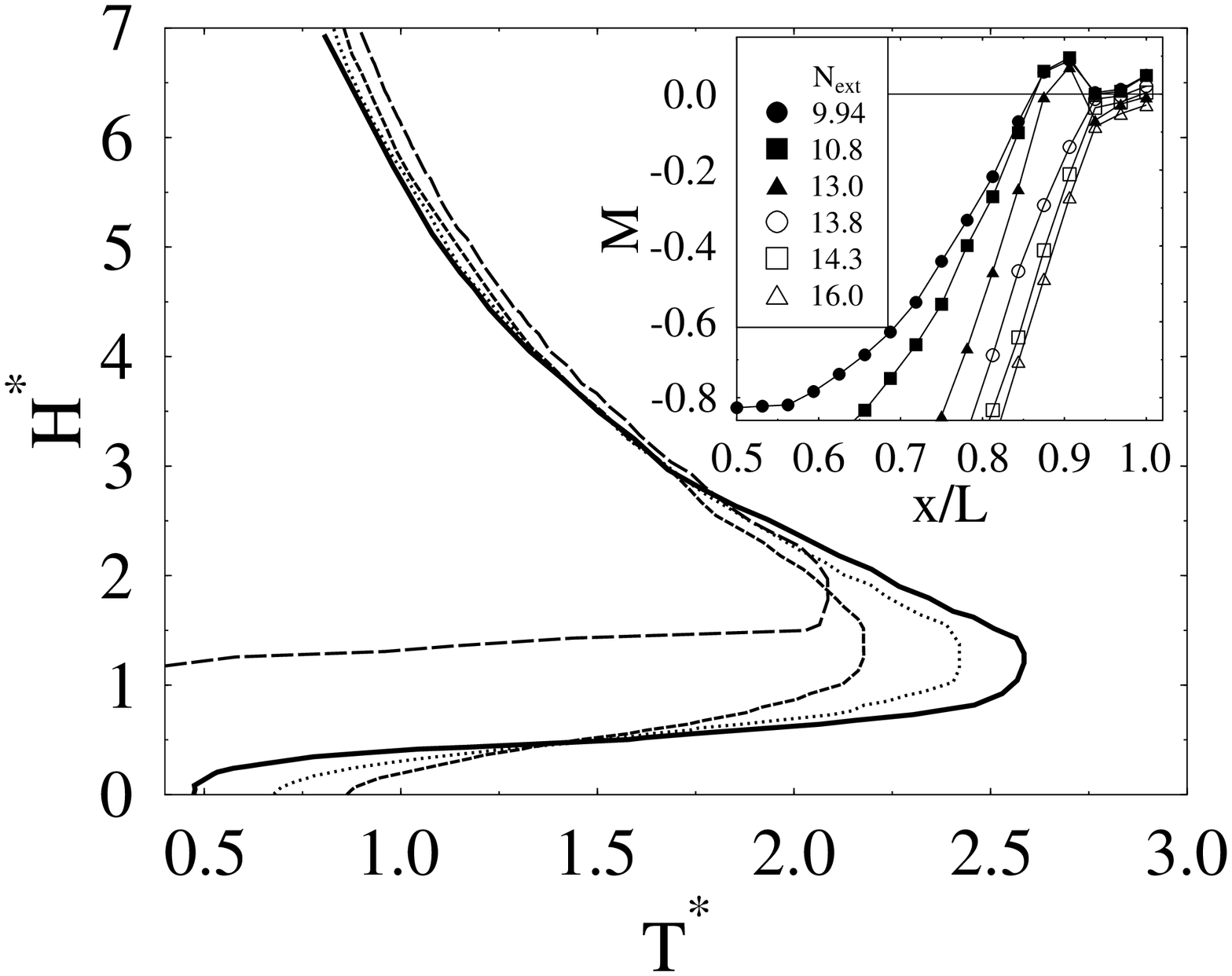,height=9.5cm,angle=-90}}
\vspace{-1.7cm}
\caption{{\bf Main frame} The mean field phase diagram of the 
ROM model in the small pinning strength regime ($A^p_0 < A_0$), 
in the plane $(H^*,T^*)$ (here $H^*=\mu/k_BT$ $T^*=T/A_1$ 
are the dimensionless chemical potential and temperature), 
for $\kappa^*=10$ and $A^p_0=0.0;0.5;0.75$ (res. full, dotted and dashed 
lines) and $\kappa^*=3.3$ and $A^p_0=0.0$ (long dashed line). 
{\bf Inset} The magnetisation profile, $M(x)$, as a function of the 
transversal spatial coordinate $x/L$ ($L$ is the system linear size), 
recorded while ramping the external field, 
$N_{ext}$ (for the shown values), in the 2D ROM model 
($\kappa^*= 0.26$, $T=0.3$, $\gamma=1.1~10^{-3}$). 
Notice the change in shapes for $N_{ext}$ smaller or 
larger than $N_p\simeq 13.5$ (filled v.s. empty symbols). 
}
\label{mfpd}
\end{figure}

To understand the equilibrium properties of the ROM model 
we briefly consider its replica mean field theory (MF). In this
approximation the equilibrium phase diagram in the plane $(H^*,T^*)$ 
(where $T^*=k_BT/A_1$ and $H^*=\mu/k_BT$) can be analytically dealt 
with (see Fig.\ref{mfpd}). 
In absence of disorder it clearly shows a reentrant phase transition from 
a high temperature low density fluid phase to an ordered phase, 
in analogy to predictions in superconductors \cite{blatter,Nelson}. 

For moderate values of the pinning energy ($A^p_0\le A_1$), a second order 
transition still takes place, which at sufficiently strong pinning  
is expected to become a ``glassy'' transition, as is seen in 
Random Field Ising Models \cite{Nattermann}. 
For the 2D lattice we consider below (in limit
$A^p\rightarrow 0$), a numerical investigation is consistent with 
a first order transition. 
In MF, the extension of the low $T$ phase shrinks 
by increasing $A^p_0$ 
(i.e., the highest critical temperature, $T_m^*$, decreases)
and the higher is $\kappa^*$ the smaller the reentrant region. 
These findings are in agreement with 
experimental results on vortex phase diagrams (see Ref.\cite{blatter} 
or, for instance, 2H-NbSe$_2$ superconductors from Ref.\cite{Bhatta}).

We now go beyond MF theory and discuss the dynamical behaviour 
of the model.
We performed MC simulations on a 2D square lattice system 
(we use typically $L^2=32^2$) 
described by eq.(\ref{H}). The system is periodic in the $y$-direction
and has the two opposite edges in the $y$-direction  
in contact with a  reservoir of particles. The reservoir is 
described by ${\cal H}$ with $A^p_i=0$ $\forall i$ and kept  
at a given density $N_{ext}$. Particles undergo diffusive dynamics 
and are introduced and escape the system only trough the reservoir.
The parameters of our simulations are usually 
$A_0=1.0$; $A_0^p=0.3$; 
$N_{c2}=27$. We have sampled several values of $\kappa^*\in[0,0.3]$. 

We are interested in the dynamical properties of the system in the 
low $T^*$ region of the above phase diagram. 
Here the 2D model has interesting magnetic hysteretic behaviours. 
In our MC simulations we ramp $N_{ext}$ (starting from zero 
and later back to zero) at a given rate 
$\gamma=\Delta N_0/\tau$ and record the magnetisation, $M=N_{in}-N_{ext}$ 
($N_{in}=\langle\sum_i s_in_i\rangle/L^d$ is the ``charge'' density inside 
the system) as a function of $N_{ext}$. Such a ramping induces a 
Bean-like profile in our lattice (inset of Fig.\ref{mfpd})
with a structure similar to some experimental data (see, for instance, 
\cite{Giller}). 

\begin{figure}[ht]
\vspace{-2cm}
\centerline{
\hspace{-2.5cm}\psfig{figure=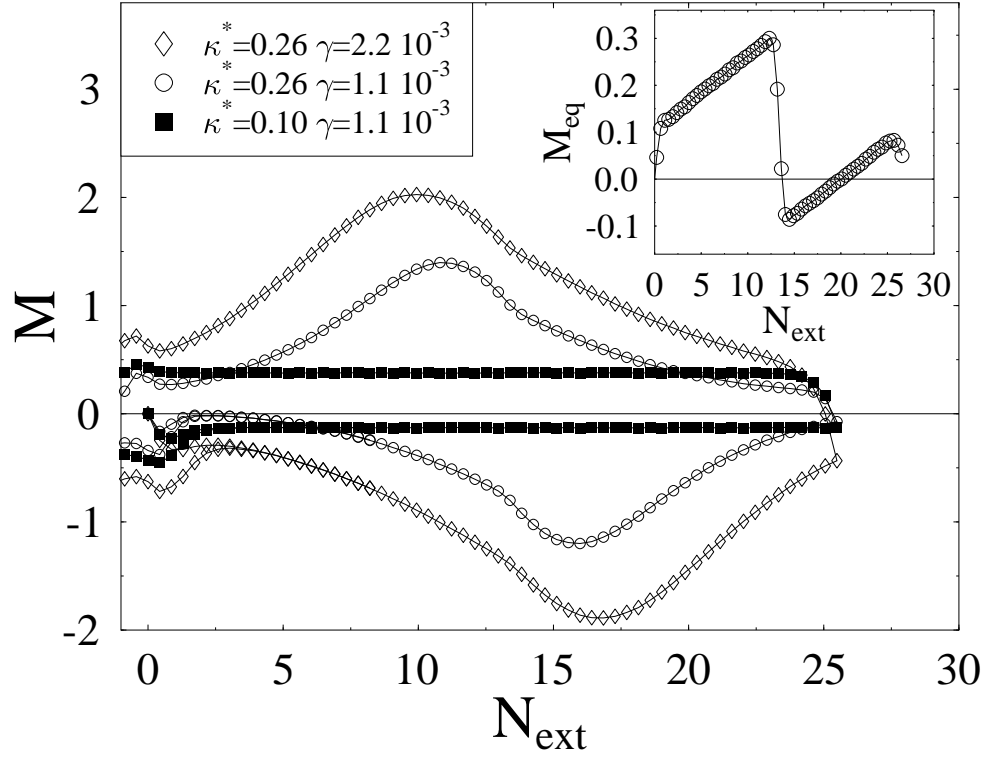,width=12cm,angle=-90} 
}
\vspace{-1.5cm}
\caption{{\bf Main frame} The magnetisation, $M$, as a function of the applied 
field density, $N_{ext}$, in the 2D ROM model for  
$\kappa^*= 0.1, 0.26$ at $T=0.3$ and the shown sweep rates $\gamma$. 
The locations of the peak, different in the increasing and decreasing 
branches, depend on $\gamma$ and approach the same value in the limit 
$\gamma\rightarrow 0$. 
{\bf Inset} 
The equilibrium value of $M$ (i.e., when $\gamma\rightarrow 0$) 
for $\kappa^*= 0.26$ at $T=0.3$. 
} 
\label{loops}
\end{figure}

\begin{figure}[ht]
$~$\vspace{-2.5cm}
\centerline{
\hspace{-2.5cm}  \psfig{figure=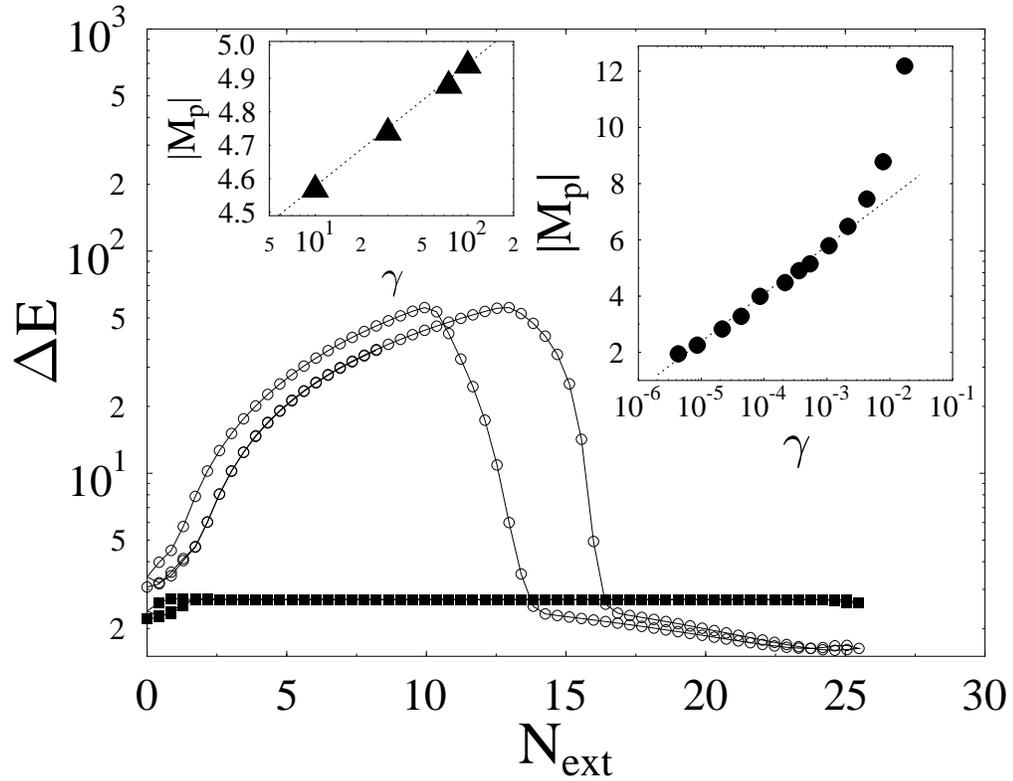,width=12cm,angle=-90} 
}
\vspace{-1.5cm}
\caption{
{\bf Main frame} The average energy barrier, $\Delta E$, a particle 
meets during diffusion in the same ROM lattices of Fig.\ref{loops}. 
{\bf Inset right} 
The second magnetisation peak height in the ROM model 
as a function of the sweep rate $\gamma$ for 
$\kappa^*=A_1/A_0= 0.28$ 
at $T=0.3$. 
{\bf Inset left} The second magnetisation peak, $M_p$ (Am$^2\times 10^3$), 
in a single crystal of YBa2Cu4O8 at temperature $20$K as a function of 
the sweep rate, $\gamma$ ($m$T/sec) (from Ref.[15]). 
} 
$~$\vspace{-0.5cm}
\label{gamma}
\end{figure}

At low temperatures ($T\le 5$ \cite{nota2}), 
a pronounced hysteretic magnetisation loop is seen 
(see Fig.~\ref{loops}), and  
when $\kappa^*$ is high enough ($\kappa^*\ge 0.25$) 
a definite second peak appears in $M$. 
In the present case the origin of the second peak is very simple. 
At high density and $\kappa^*$, 
groups of vortices, frustrated in minimising their 
repulsive interaction energy, are forced to cluster together 
forming macroscopically extended energetic barriers which cage other diffusing 
vortices. 
In Fig.\ref{gamma} we plot the average energy barrier, $\Delta E(N_{ext})$, 
a particle meets during the same runs for $M$ shown in Fig.\ref{loops}. 
A ``trial'' vortex approaching groups of 
clustered vortices has to pass over these barriers to move further. 
This dynamically generates the second peak. 
The final decrease in $M$ at high $N_{ext}$ is, here, due to 
a ``softening'' of these barriers caused by saturation effects 
related to the finite value of $N_{c2}$. 
The first peak in the magnetisation stems from the fact that density 
variations 
in the reservoir are only slowly transmitted in the system when it is in 
the low density ``fluid'' phase. 
The second peak and hysteretic loops at moderate-high 
$\kappa^*$ are also present when $A^p_0\rightarrow 0$ ($A^p_0$ also determines
the difference in the amplitude of $|M|$ 
between the increasing and decreasing ramps). 
Very similar magnetisation data are observed in a number of different 
superconductors from intermediate to high $\kappa$ values
(see, for instance, ref.s in \cite{cohen,Bhatta,Paltiel}).

The actual shape of loops strongly depends on the 
parameters of the dynamics (and system size). In particular, 
the sweep rate of the external field, 
$\gamma$, is very important. 
As soon as the inverse sweep rate 
is smaller than the characteristic relaxation time 
(which can be extremely long, inaccessible on usual observation time scales, 
see below) strong off-equilibrium effects are present, 
such as metastability or ``memory'' and ``aging'' 
\cite{Yeshurun,Bhatta,Paltiel}. 
As a first example of these facts, we show in the right 
inset of Fig.\ref{gamma} the dependence of the second peak 
height, $M_p$, on $\gamma$. At low temperatures $T\leq 1$ and not too 
low $\kappa^*$ ($\kappa^*\ge 0.28$), $M_p$ is approximately logarithmically 
dependent on $\gamma$ over several decades: 
\begin{equation}
M_p(\gamma)\simeq M_0+\Delta M \ln(\gamma)
\end{equation}
Such a behaviour gets closer to a power law $M_p\sim \gamma^x$ 
at lower $\kappa^*$ (for instance, $x\sim 1/2$ at $\kappa^*=0.26$). 
Eventually, when $\gamma$ is smaller 
than a characteristic threshold, $\gamma_t$, 
$M_p$ exponentially saturates to its asymptotic value (this usually is 
orders of magnitude smaller than $M_p(\gamma)$ at high $\gamma$). 
Interestingly these findings are also very close to what is experimentally 
observed in superconductors \cite{cohen,Yeshurun}; an example 
from an YBCO sample (from \cite{Perkins}) 
is given in the left inset of Fig.\ref{gamma}. 
The threshold, $\gamma_t$, is 
strongly dependent on the system density $N_{in}$ (and system size) and is 
a rapidly decreasing function of $\kappa^*$;  for instance at $T=0.3$, 
for $N_{ext}=N_p$ ($N_p$ is the location of $M_p$ ), 
$\gamma_t|_{\kappa^*=0.26}\simeq 4.5\cdot 10^{-5}$ but 
$\gamma_t|_{\kappa^*=0.28}\leq 10^{-6}$. 
$\gamma_t^{-1}(T,N_{in};\kappa^*,L)$ 
is a measure of the system characteristic equilibration times (which can 
be  huge). 

Seemingly a dynamical phenomenon, 
in the ROM model the second peak is related to a true transition: 
in the $\gamma\rightarrow0$ limit, its location, $N_p$, 
is associated with a sharp jump in 
$M_{eq}\equiv\lim_{\gamma\rightarrow 0} M(\gamma)$, where its fluctuations 
increase with system size (see inset of Fig.\ref{loops}). 
These findings are consistent with experiments 
(for instance, see Ref.\cite{Bhatta}) and 
to some extents reconcile opposite descriptions 
(``static'' v.s. ``dynamic'') of the phenomenon. 

\begin{figure}[ht]
$~$\vspace{-2.5cm}
\centerline{
\hspace{-2.5cm}  \psfig{figure=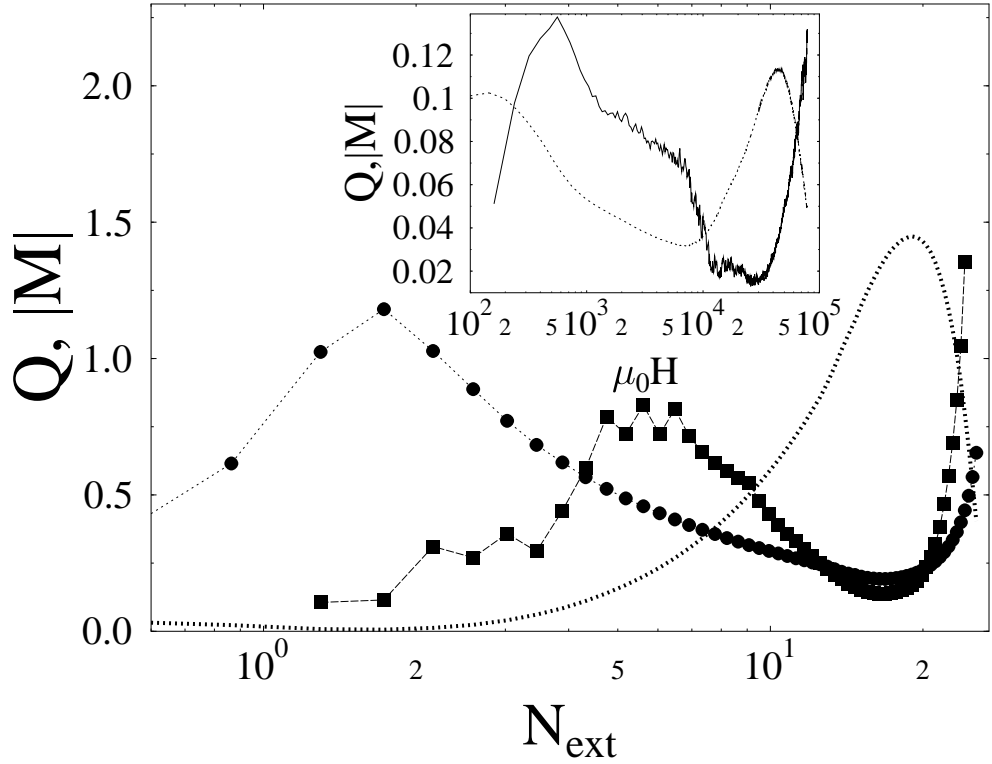,width=12cm,angle=-90} 
}
\vspace{-1.5cm}
\caption{{\bf Main frame} For the same ROM models of the right inset of 
Fig.\ref{gamma} with $\kappa^*=0.28$,  
the ``creep rate'' $Q\equiv {\partial \ln M \over \partial \ln \gamma}$ 
averaged over the intervals $\gamma\in[5\cdot 10^{-3}, 10^{-1}]$ 
(filled circles) 
and $\gamma\in[5\cdot 10^{-4}, 5\cdot 10^{-3}]$ (filled squares), 
is  plotted as a function of $N_{ext}$. For comparison, 
a corresponding magnetisation loop ($M\rightarrow M/4$ to have clear scale 
on y-axis) is also shown (dotted line). 
{\bf Inset} The creep rate (for $\gamma\in[10,100]$$m$T/sec, full line) 
and magnetisation loop ($4\cdot 10^2$Am$^2$, dotted line) as a function of 
the external magnetic field $\mu_0H$ (T) 
in the same YBa2Cu4O8 sample of Fig.\ref{gamma}. 
} 
$~$\vspace{-0.5cm}
\label{Q}
\end{figure}

It is also interesting to consider the ``creep rate'' 
$Q={\partial \ln M \over \partial \ln \gamma}$, which is often 
associated to a measure of the intrinsic energy barriers in the 
creep process \cite{Yeshurun}. 
Experimentally, $Q$ is a non trivial function of the 
magnetic field (see for instance \cite{Perkins,Yeshurun}). 
We find that, due to the very long relaxation times, $Q(N_{ext})$ 
is in itself a (slowly varying) function of $\gamma$, up to when 
$\gamma$ is smaller than the smallest $\gamma_t$. 
In Fig.\ref{Q} we show how $Q$ depends on $\gamma$ in the ROM model: 
for $T=0.3$ and $\kappa^*=0.28$ 
we plot as a function of $N_{ext}$ the average of $Q$ over two different 
$\gamma$ intervals $\gamma\in[5\cdot 10^{-3}, 10^{-1}]$ (filled circles) 
and $\gamma\in[5\cdot 10^{-4}, 5\cdot 10^{-3}]$ (filled squares). 
The difference between the two is apparent. 
We note a remarkable correspondence with experimental data 
for YBCO, shown for the quoted sample in the inset of Fig.\ref{Q}. 

\begin{figure}[ht]
\vspace{-1.6cm}
\centerline{
 \hspace{-2.5cm} 
 \psfig{figure=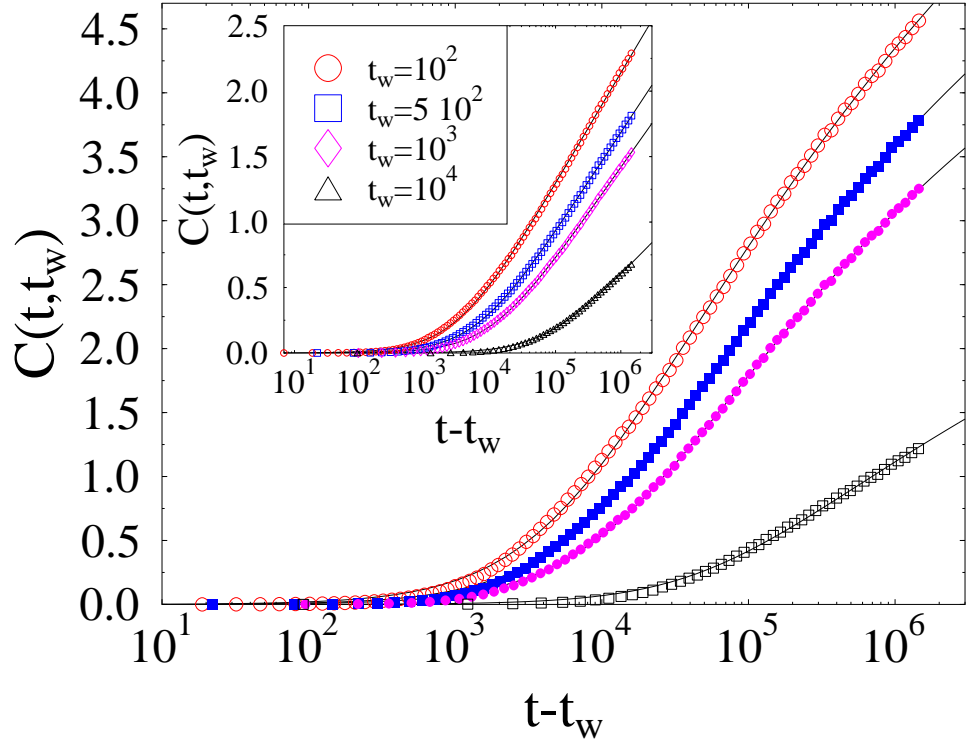,width=11.5cm,angle=-90}
}
\vspace{-1.8cm}
\caption{
Time relaxation of the two-times vortex-density 
correlation  function, $C(t,t_w)$, in the 2D ROM model, recorded at $T=0.1$ 
($\kappa^*=0.28$) for the shown $t_w$ at $N_{ext}=4,16$ 
(resp. inset, main frame). Continuous lines are logarithmic fits.} 
\label{rel}
\end{figure}

The presence of the above ``memory'' effects 
indicate that the system, on the observed 
time scales, can be well off equilibrium. 
To reveal the underlying non-stationarity of the dynamics
we consider two times correlation functions and, at a given $N_{ext}$, 
we record ($t>t_w$) \cite{nota3}: 
\begin{equation}
C(t,t_w)=\langle [N_{in}(t)-N_{in}(t_w)]^2\rangle ~ .
\label{cor}
\end{equation}
Fig.\ref{rel} clearly shows that $C(t,t_w)$ exhibits strong ``aging'':
it explicitly depends on both times, 
contrary to situations close to equilibrium, where $C$ is 
a function of the times difference $t-t_w$. 
In particular at high $\kappa^*$ and low $T$ 
(where relaxation times are very high) $C(t,t_w)$ can be well 
fitted by a generalisation of a known interpolation formula, 
often experimentally used \cite{blatter}, which now depends on 
the {\em waiting time}, $t_w$: 
$C(t,t_w)\simeq C_{\infty}\left\{1-\left[1+
\frac{\mu T}{U_c}\ln\left(\frac{t+t_0}{t_w+t_0}\right)\right]^{-1/\mu}\right\} 
$.
We found that to take $\mu\simeq 1$ is consistent with our data. Notice 
the presence of {\em scaling properties}: for not too short times 
$C$ is a function of only the ratio $t/t_w$: 
$C(t,t_w)\sim {\cal S}(t/t_w)$. This is a fact in agreement with
general scaling in off equilibrium dynamics (see  Ref.\cite{CoNi}) and 
in strong analogy with other systems (from glass formers 
to granular media \cite{BCKM,CoNi,Angell,NC_aging}). Experimental measurements
of $C(t,t_w)$ would be {\em very} valuable.

In conclusion, in the context of a simple tractable model 
we depicted a panorama of magnetic properties of 
vortices in type-II superconductors. 
Even the 2D version of the model has many interesting features 
in correspondences with experimental results 
and allows clear predictions on the nature vortex dynamics. 
The origin of the slow off equilibrium relaxation (observed at low $T$) 
is caused by the presence of very high free energy barriers 
{\em self generated} by the strong repulsive interaction 
between particles at high densities (for $\kappa^*$ above a threshold). 
In this respect the pinning potential plays a minor role. 
For instance, the presence of a ``second peak'' in $M$, is also observed 
in the limit $A^p\rightarrow 0$.
The second peak doesn't mark the transition to a ``glassy'' 
phase, but is also present in such a case. $A^p$ 
sets the position and amplitude of the reentrant order-disorder 
transition line, which is in turn distinct from the second peak locations. 

At low temperatures on typical 
observation time scales, the system is strongly off equilibrium. 
This is most clearly seen from ``aging'' found 
in two-times correlation functions. These obey scaling properties of purely 
dynamical origin 
Experimental check of these results would be extremely important 
to elucidate the true nature of vortex dynamics.

{\bf Acknowledgements}   
We thank L. Cohen and G. Perkins for useful discussions 
and for the YBCO data. Work supported by the EPSRC 
and PRA-INFM-99.

\newpage

\newpage

\newpage

\newpage

\begin{thebibliography}{40}
\bibitem{nota1} 
Asymptotically, vortex line segments interaction 
is exponential $V(r)\sim V_0\exp(-r/\sqrt{2}\lambda)$ 
\cite{Brandt}. One can write: $A_0/A_1\propto V(0)/V(const. \xi)$ 
$\Longrightarrow$ $1/\kappa^*\propto\exp(const. /\kappa)$.
\bibitem{nota2} We use $A_0=1.0$ as the energy unit and set $k_B=1$.
\bibitem{nota3} Time is measured in units of a MC full lattice sweep.
\end{thebibliography}
\end{document}